\def\spose#1{\hbox to 0pt{#1\hss}}
\def\lta{\mathrel{\spose{\lower 3pt\hbox{$\sim$}}
    \raise 2.0pt\hbox{$<$}}}
\def\gta{\mathrel{\spose{\lower 3pt\hbox{$\sim$}}
    \raise 2.0pt\hbox{$>$}}}
\documentclass[12pt,preprint]{aastex}

\usepackage{amsmath}
\usepackage{graphicx}
\received{2003 January 24}
\begin{document}

\title{A dynamical fossil in the Ursa Minor dwarf spheroidal galaxy}
\author{Jan T. Kleyna$^1$, Mark I. Wilkinson$^2$,  Gerard Gilmore$^3$, 
N. Wyn Evans$^4$,}
\affil{$^{1,2,3,4}$Institute of Astronomy, Madingley Road,Cambridge, CB3 OHA, UK}
\affil{$^1$kleyna@ast.cam.ac.uk; $^2$markw@ast.cam.ac.uk; 
$^3$gil@ast.cam.ac.uk; $^4$nwe@ast.cam.ac.uk }

\begin{abstract} 
  The nearby Ursa Minor dwarf spheroidal (UMi dSph) is one of the most
  dark matter dominated galaxies known, with a central mass to light
  ratio $M/L\sim70$.  Somewhat anomalously, it appears to contain
  morphological substructure in the form of a second peak in the
  stellar number density.  It is often argued that this substructure
  must be transient because it could not survive for the $>10$ Gyr age
  of the system, given the crossing time implied by UMi's 8.8 $\rm
  km\,s^{-1}$ internal velocity dispersion.  In this paper, however,
  we present evidence that the substructure has a cold kinematical
  signature, and argue that UMi's clumpiness could indeed be a
  primordial artefact. Using numerical simulations, we demonstrate
  that substructure is incompatible with the cusped dark matter haloes
  predicted by the prevailing Cold Dark Matter (CDM) paradigm, but is
  consistent with an unbound stellar cluster sloshing back and forth
  within the nearly harmonic potential of a cored dark matter halo.
  Thus CDM appears to disagree with observation at the least
  massive, most dark matter dominated end of the galaxy mass spectrum.
\end{abstract}


\section{Introduction}
The CDM model of structure formation \citep[e.g.][]{ostriker93}
postulates that galaxies form when initial perturbations in a sea of
cold (non-relativistic) dark matter particles seed the growth of
larger dark matter haloes characterised by a central density cusp
$\rho(r)\propto r^{-a}$, where $a=1$ to $1.5$ \citep{navarro97, ghigna00}.  However, the rotation curves of massive galaxies almost
invariably suggest that haloes have a flat density core rather than a
sharp cusp \citep[e.g.][]{deblok01}. It has been suggested that an
initial cusp may be destroyed by feedback mechanisms that couple
supernovae driven winds to the dark matter \citep[e.g.][]{binney01}, but
models of this sort have had mixed success.

The dSph galaxies \citep[e.g.][]{mateo98} surrounding the Milky Way
provide a good laboratory in which to test CDM predictions on small
scales: most are sufficiently nearby to allow us to measure the
velocities of hundreds of member stars, and most have such meagre
baryonic content that their internal dynamics is everywhere dominated
by the dark matter.  Although it has argued for some time that CDM
greatly overpredicts the number of dwarfs, some recent work
\citep{stoehr02} suggests that CDM is actually in good agreement with
the local dSph population.  Ursa Minor is, with Draco, one of the two
most dark matter dominated dSphs of the Local Group, with a central
mass to light ratio $M/L\sim 70 M_\odot/L_\odot$ \citep{HargreavesUMi,
  arm95}.  UMi's stars appear to have been formed in a single burst
\citep{carrera02}, and its size, stellar velocity dispersion, and
luminosity resemble those of Draco, which has been shown to have a
{\sl mean} mass to light ratio $M/L\sim400 M_\odot/L_\odot$ within
three core radiii \citep{kleyna01}.  Unlike most other dSphs, however,
UMi has substantial morphological distortions: UMi is highly
elongated, and appears to possess a secondary clump or shoulder on the
northeast side of the major axis \citep{IrHatz95, kleyna98, PalmaUMi}.
It is often argued that this clump cannot be a persistent feature
because the $\sim 2\times10^7$ year stellar crossing time of the
system is orders of magnitude shorter than its $\sim 10^{10}$ year
age, so that stellar orbits within an unbound clump will diverge over
hundreds of dSph crossings.  UMi's relatively circular orbit
\citep{schweitzer03} weighs against the common explanation
that the clump is a temporary artefact generated through ongoing tidal
disruption by the Galaxy.  Generally, models that attempt to explain
the dSphs' large velocity dispersions by tidal disruption still
require dark matter, or else must postulate an unseen supply of dSphs
to replace those that are disrupted \citep{oh95}, or require that
dSphs are disintegrated remnants viewed along a very fortuitous line
of sight \citep{kroupa97}.

In this paper, we use a data set consisting of new and extant UMi
stellar velocities to argue that the second peak in UMi's stellar
distribution has a cold kinematical signature, and we demonstrate that
a plausible explanation is that the clump is a disrupted stellar
cluster sloshing back and forth within UMi's halo.  We use numerical
simulations and analytical arguments to show that persistent
substructure is consistent with a cored halo, but not with a cusped
CDM halo.

\section{OBSERVATIONS AND ANALYSIS}

In May 2002, we obtained spectra of 63 stars in UMi using the WYFFOS
multifibre spectrograph at the William Herschel Telescope on La
Palma. We cross-correlated the spectra with a synthetic template of
the Ca triplet lines at $\sim 850$ nm to measure velocities
\cite[e.g.][]{kleyna02}; the median velocity error was 5 $\rm km\,s^{-1}$,
less than the UMi's 8.8 $\rm km\,s^{-1}$ central velocity dispersion.
As our stars were mostly at large projected radii in UMi, we combined
our data with the more centrally concentrated sample of Armandroff et
al. (1995), after subtracting the mean velocity difference of the two
data sets.  The combined data set contains 134 stars.  Because the
observing run was curtailed by bad weather, the data did not extend to
a sufficiently large radius to allow a fit to the halo shape and
anisotropy \citep[c.f.][]{wilkinson02, kleyna02}.  However, we noted that a
histogram of stellar velocities near the clump on the northeast side
of UMi's major axis appeared narrower than the $8.8\,\rm km\,s^{-1}$
Gaussian describing UMi's overall line of sight velocity distribution
(Figure 1). This observation suggested constructing a model of UMi's
population consisting of the sum of two Gaussians, one representing
the underlying $\sigma_0=8.8\,\rm km\,s^{-1}$ Gaussian, and the other
representing a sub-population of unknown fraction $f$, velocity offset
$v_s$, and velocity dispersion $\sigma_s$.  Assuming an observational
velocity uncertainty $\sigma_{\rm obs}$, the probability of observing
a particular velocity in this model is

\begin{eqnarray}
P(v|f,\sigma_s,\sigma_0,\sigma_{\rm obs})& = &
f \times 
{\exp {\left[-{1\over 2} {(v-v_s)^2\over {\sigma_s}^2 + 
  \sigma_{\rm obs}^2}  \right]}
 \over \sqrt{2\pi ({\sigma_s}^2 + \sigma_{\rm obs}^2)}
 }\,\, +\\
& &
(1-f)\times 
{\exp {\left[-{1\over 2} {v^2\over \sigma_0^2 + \sigma_{\rm obs}^2}  \right]}
 \over \sqrt{2\pi (\sigma_0^2 + \sigma_{\rm obs}^2)}
 } \nonumber 
\end{eqnarray}
\noindent and the likelihood of obtaining a ensemble of velocities
$\{v_i\}$ with errors $\{{\sigma_{\rm obs}}_i\}$ is 
\begin{equation}
 P(\{v_i,{\sigma_{\rm obs}}_i \})= 
\prod_i P(v_i|f,\sigma_s,\sigma_0,{\sigma_{\rm obs}}_i)
\end{equation}
Next we scanned the face of UMi in RA and Dec in $2^\prime$
increments.  At each point, we collected all velocities in a
$6^\prime$ radius aperture, and computed the statistical likelihood of
drawing these velocities from an 8.8 $\rm km\,s^{-1}$ Gaussian
(Equation 2 with $f=0$), as well as the likelihood of drawing the
velocities from each member of a grid of two-Guassian models (Equation
2, with $f=0.1, 0.2 \ldots 1.0$, $v_s=-10, -9 \dots 10$,
$\sigma_s=0.5,1.0\ldots 15.0$).  When the likelihood of the
best-fitting two-Gaussian model exceeded the likelihood of the single
Gaussian model by a large factor, we deemed that aperture to contain a
kinematic subpopulation distinct from the rest of UMi's stars.  Figure
2 shows the result of the scanning procedure.  Large dots are
apertures where we found a significant (likelihood ratio $>10^3$)
sub-population.  At the largest dot, the best two-Gaussian population
( $\sigma_s=\,0.5 \rm km\,s^{-1}$, $v_s=\,-1\rm km\,s^{-1}$, $f=0.7$)
is $4.7\times10^4$ times more likely than the default 8.8 $\rm
km\,s^{-1}$ single-Gaussian model.  Moreover, the largest dot is very
close to where the isopleth contours indicate the presence of a
secondary peak in the stars.  The small value of $v_s$ suggests that
the sub-population is either on a radial orbit or a face-on circular
orbit. We note that our best-fit $\sigma_s$ is ill-determined because
it is much smaller than the velocity measurement errors .  Also, this
small value of $\sigma_s$ applies only to the surviving phase space
region of the progenitor cluster - any faster stars may have joined
UMi's general population.  We verified the statistical significance of
the second kinematical population with a Monte-Carlo test: we
repeatedly generated artificial data drawn from an 8.8 k m/s Gaussian
at the same locations and with the same velocity uncertainties as our
actual data, and subjected it to the scanning procedure described
above.  We found that only 11 of 2000 (0.055\%) artificial data sets
give as large a likelihood ratio as the actual data anywhere in UMi.
Only one of these false positives suggested an orbit in the plane
of the sky, and all of the others had $v_s>5$.

\section{IMPLICATIONS OF THE COLD SUB-POPULATION}

To investigate the survival of a cold kinematical sub-population in a
dSph, we integrated the motion of a group of unbound stars inside a
plausible halo model.  We assumed a halo density law
$\rho(r)\propto(a^2+r^2)^{-1/2}$, which has a core for $a>0$ and a
$r^{-1}$ cusp for $a=0$.  We dropped an unbound clump modelled as a
three dimensional spatial Gaussian with a three dimensional Gaussian
velocity dispersion into this halo, and followed its evolution with an
adaptive stepsize Runge-Kutta code for 12 Gyr.  Using the mass of
Draco as a reference
\citep{kleyna01}, we normalised the enclosed mass of the model to 
$5\times10^7 M_\odot$ inside $r=1$, where $r=1$ corresponds to 600 pc,
approximately the maximum extent of the stellar distribution.  In our
dimensionless units, the clump orbits within $r=0.25$, or 150 pc
in physical units.

Figure 3 shows the results of the simulations.  When a clump with an
initial one dimensional dispersion $\sigma_s = 0.5 \,\rm km\,s^{-1}$
and a one-sigma radius of 0.02 (12 pc) is dropped at an initial radius
$r_0=0.25$ into a halo with an $a=0.85$ core (top panel) on
near-radial orbit with a tangential velocity equal to 0.07 times the
circular speed at $r_0$, the resulting sub-structure is visible for
many Gyr.  Even at 12 Gyr, nearly the age of the Universe, the
remnants of the clump are spatially concentrated and asymmetrical.
When an identical clump is dropped into a $a=0$ cusped halo (bottom
panel) the clump dissipates completely within 1 Gyr.

The total velocity $v=(v_x^2+v_y^2+v_z^2)^{1/2}$ histograms of Figure
3 demonstrate that the stars remain cold for much of the evolution of the
cored model, whereas they quickly achieve a flat distribution in a
cusp.  However, the fact that the orbit is constructed to be
approximately in the plane of the sky implies that the measured
line-of-sight $v_z$ distribution remains cold for both the core and
cusp.  The distinguishing observable feature of the cored model,
therefore, is a {\sl localized} population that is cold in the
line-of-sight.

Further simulations demonstrate that this conclusion holds good both
for more circular orbits and for triaxial haloes.  Cusps shallower
than $r^{-1}$ only slightly increase the longevity of the clump: for
example, a clump in a $\rho(r)\propto r^{-0.5}$ cusp survives
approximately twice as long as in a $r^{-1}$ cusp.  If the halo core
is an order of magnitude larger than the clump's orbit, a clump can
survive almost unalterned for a Hubble time.  If the core radius is
equal to or smaller than the clump's orbit, however, the clump is
destroyed within one or two Gyr.  We conclude that a clump can survive
for a large fraction of a Hubble time only if it is embedded in a core
at least two or three times the size of its orbit.

An additional concern is the possibility that Galactic tides may
destroy substructure in a cored halo.  To address this point, we added
the effect of tides to our simulation. We modelled the Galaxy as an
isothermal sphere of mass $10^{12}\,M_\odot$ inside 50 kpc, somewhat
more massive than the value of \cite{WE99}; at a 60 kpc
apogalacticon, the tidal force from this Galaxy model is 0.02 times as
strong as the dwarf's gravity at a unitless dSph radius $r=1$.  We
then created a varying tidal force in the dwarf's $x,y$ plane by
moving the isothermal Galaxy model around the dSph on 1.6 Gyr orbits
with eccentricities between 0 and 0.5.  In no instance did this weak
tidal force significantly affect the survival of substructure either
in a core or in a cusp.  Indeed, we could safely make the tidal force
an order of magnitude larger without causing the clump to
disintegrate in the above  $a=0.85$ core.

We emphasize that these simulations are not in any sense a fit to the
UMi data.  Rather, they demonstrate that substructure can easily
persist for many Gyr in a plausible cored halo, but is quickly erased
in a cusp.  Several unconstrained parameters affect the survival of
substructure in a core.  In addition to increasing the core size, for
example, one can decrease the dSph's mass, slowing the internal time
scale and decreasing the number of orbits over which a clump can
dissipate.  Similarly, one can increase the longevity of the clump by
decreasing its initial size or velocity dispersion.  We find that
strong substructure continues to persist for 12 Gyr when we double or
treble the size or dispersion of our model clump in Figure 3, although
the final distribution is less compact than in our illustrative
example.

We note that there is a simple explanation for the persistence of a
clump in any cored halo: a generic feature of dark matter haloes
with cores is that the potential is close to that of a harmonic
oscillator.  The one dimensional $x$ motion of a particle with
position $x_1$ and velocity $v_1$ at time $t=0$ is $X_1(t) = x_1 \cos
(\omega_x t) + v_1 \omega_x^{-1} \sin(\omega_x t)$.  Then the $x$
distance between two particles dropped into a harmonic potential at
positions and $x_1, x_2$ with velocities $0,v_2$ is given for all time
as
\begin{equation}
X_1(t)-X_2(t) = (x_1-x_2)\cos (\omega_x t) + 
x_2 \left({v_2 \over {x_2 \omega_x}}\right) \sin (\omega_x t)
\end{equation}

The cosine term is bounded by the initial separation, and the sine
term is bounded by the orbital amplitude times the ratio of the
relative speed (dispersion) to the orbital speed, in the approximation
that the dispersion is smaller than the orbital speed.  Because the
three dimensions of the harmonic oscillator are independent, this
result does not depend on spherical symmetry and is valid for any
triaxial harmonic potential.  Hence, if the non-harmonic terms of the
Taylor expansion of the potential are sufficiently small, an unbound
stellar clump with a small but finite internal dispersion can persist
over a long time, pulsating in size without disintegrating.  

In view of the above discussion, the fact that the tidal force only
weakly affects clump survival is not surprising. The lowest order term
of the tidal force arises from a rotating quadratic effective
potential. This quadratic potential changes adiabatically, because the
dwarf's orbit around the Galaxy is $\sim 10^2$ times longer than the
internal time scale.  Thus particles on adjacent sinusoidal orbits
remain on adjacent sinusoidal orbits as the potential evolves.

The low-dispersion substructure that we observe in UMi is consistent
with a clump progenitor similar to star clusters seen elsewhere in the
Local Group.  By rescaling the number of stars in the best aperture by
the best-fit subpopulation fraction $f=0.7$, we estimate that the
clump contains $\sim7\%$ of UMi's population.  If we assume that UMi
has a normal stellar mass to light ratio $2 M_\odot/L_\odot$ and a
total V-band luminosity $L_v=\sim2\times10^5 L_\odot$, the clump has a
luminosity of $1.5\times10^4$ $L_\odot$ and a mass $M=3\times10^4
M_\odot$.  These parameters are similar to the low mass globular
clusters in the SMC and LMC.  A cluster with the mass of our clump and
a half-mass radius $r_{1/2}=10\,\rm pc$ has a characteristic
one-dimensional internal velocity $\sqrt{1/3} \sqrt{G (M/2)/r_{1/2}
  }\approx 1.5 \, \rm km\,s^{-1}$, cold compared to
UMi's overall population.  Such low-mass clusters can become unbound
immediately after formation when supernovae expel the gas content
\citep{goodwin97}, or possibly slowly though tidal interactions with
the parent galaxy's halo \citep{giersz97}.  

\section{CONCLUSIONS}

Ursa Minor is known to possess a second peak in its stellar
distribution.  We demonstrate that the stars in the vicinity of this
peak comprise a kinematically distinct cold sub-population; the
probability of obtaining a similarly strong false positive in the
absence of a sub-population is $0.055\%$.  We show that the properties
of this clump are consistent with the remnants of a disrupted stellar
cluster orbiting in the plane of the sky within the harmonic potential
of cored dark matter halo.  Although a clump can persist for a Hubble
time in a harmonic potential, the potential of a CDM cusped halo would
destroy the clump within $<1$ Gyr.  This finding suggests that the
haloes of even the least massive, most dark matter dominated galaxies
possess a core rather than a cusp.

\acknowledgments 

\noindent NWE is supported by the Royal Society, while MIW and JK acknowledge
support from PPARC.  The authors gratefully thank the staff of the Isaac
Newton Group for help in acquiring the data, and the anonymous
referee for useful criticisms and suggestions.




\clearpage
\begin{figure}
\epsscale{0.6}
\plotone{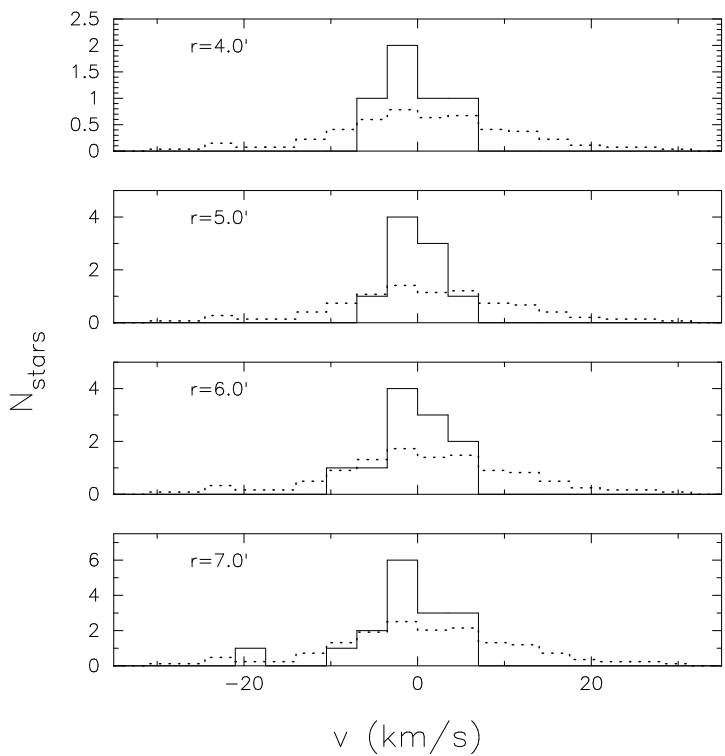}
\vskip 1cm
\caption{Solid histogram is UMi's stellar velocity distribution
inside apertures with diameters $4^\prime$ to $7^\prime$, centered on
the location of second clump (${\rm RA}=12^\prime, {\rm Dec}=8^\prime$
relative to UMi's center; see Fig. 2).  The dotted histogram is the
distribution of velocities of the entire UMi sample, normalized to
have the same number of stars as the solid histogram.  }
\end{figure}

\clearpage

\begin{figure}
\epsscale{0.75}
\plotone{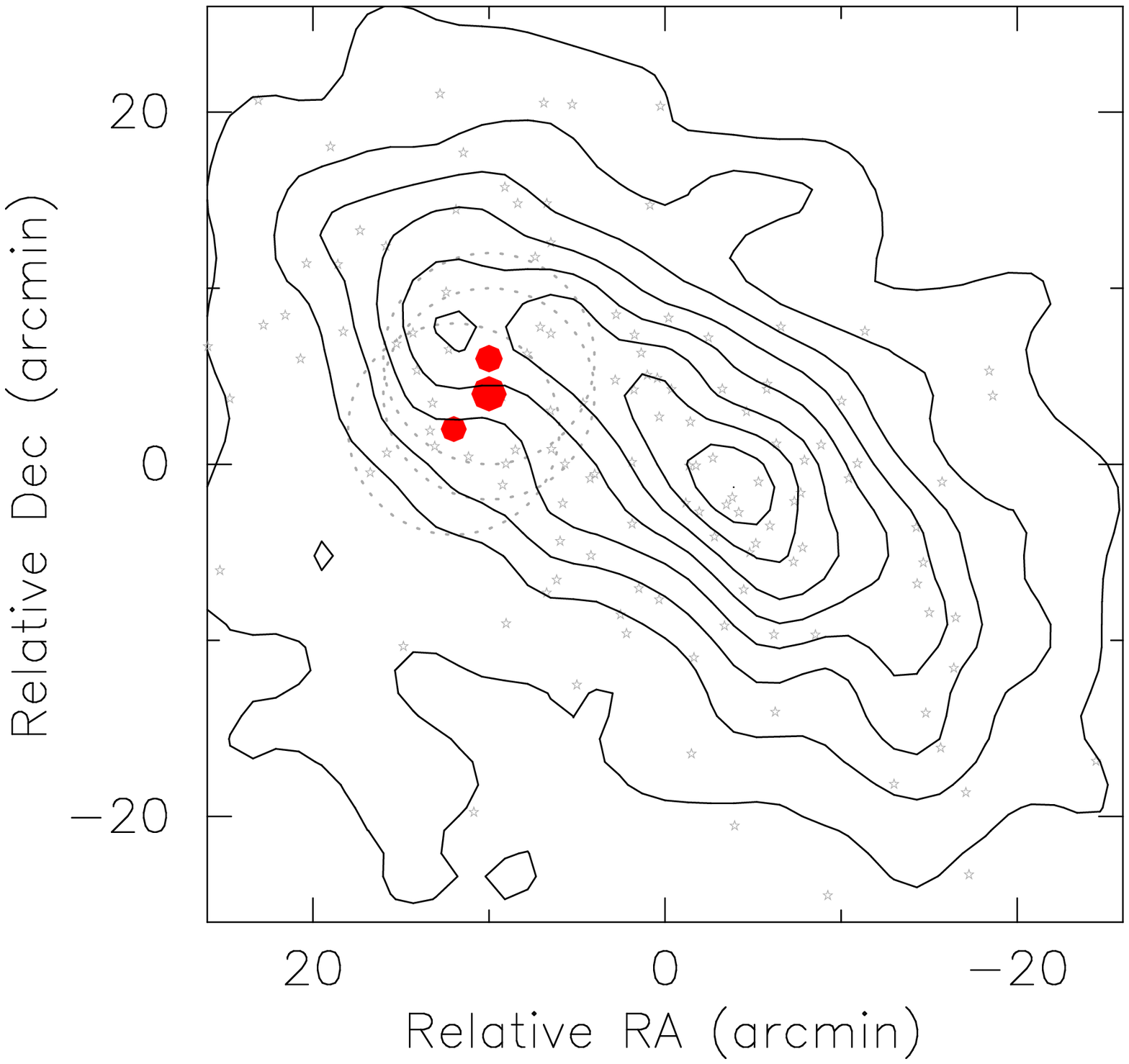}
\vskip 1cm
\caption{Result of search for kinematic sub-populations in UMi. Coutours
are linearly spaced stellar isopleths; the second peak of UMi's
stellar population is visible at ${\rm RA}=12^\prime, {\rm
Dec}=8^\prime$ relative to UMi's center at $\rm
15^h09^m10^s\hskip-3pt.2$, $\rm +67^\circ12^\prime52^{\prime\prime}$
J2000.  Gray stars are UMi RGB member stars with measured
velocities. The filled circles represent points where a model with a
kinematically cold sub-population is at least 1000 times more likely
than a model composed of a single $8.8\,\rm km\,s^{-1}$ Gaussian.  The
dotted circles are the apertures containing the stellar sample for
each dot.  The size of each dot is proportional to the logarithm of
the likelihood, and the largest dot represents a likelihood ratio of
$4.7\times10^4$, with an optimal clump fraction $f=0.7$, mean clump
velocity $v_s=-1\,\rm km\,s^{-1}$, and clump dispersion $\sigma_s
=0.5\,\rm km\,s^{-1}$. }
\end{figure}
\clearpage

\begin{figure}
\epsscale{0.9}
\plotone{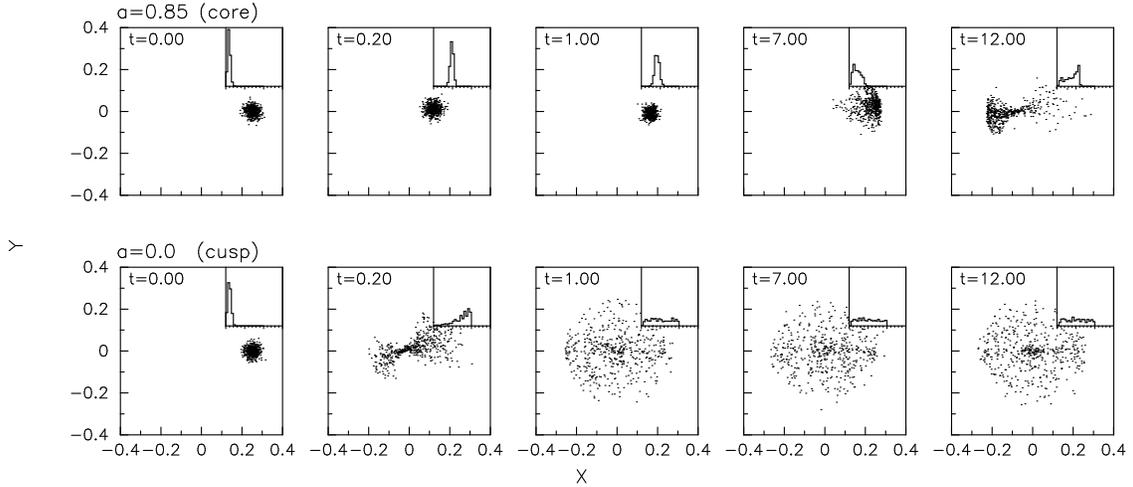}
\vskip 1cm
\caption{Simulation of an unbound clump in a dark matter halo.
The halo has a density law $\rho(r)\propto(a^2+r^2)^{-1/2}$ and a mass
of $5\times10^7 M_\odot$ inside $r=1$, which corresponds to 600 pc. The clump
has an initial one-dimensional dispersion of $0.5\,\rm km\,s^{-1}$ and
an initial speed equal to 0.05 times the circular speed in each of the
the $y$ and $z$ directions; the time $t$ is in units of
Gyr. Top panel: a clump in a cored halo with $a=0.85$ persists for a
Hubble time because the potential is nearly harmonic.  Bottom panel: a
clump in a cusped potential ($a=0$) disrupts in less than 1 Gyr.
The histograms at the upper right corner of each snapshot show
the distribution of total velocity $v=(v_x^2+v_y^2+v_z^2)^{1/2}$;
tick marks are spaced 1 $\rm km\,s^{-2}$ apart.
As is expected, the stars in a cored halo remain coherent in velocity
as well as in position.}
\end{figure}
\clearpage

\end{document}